# Conversation Networks

Deb Roy[1], Lawrence Lessig[2], Audrey Tang[3]
Feb 25, 2025

## Meeting the Needs of Communities

Picture a community torn over a proposed zoning law. Some are angry, others defensive, and misunderstandings abound. On social media, they broadcast insults at one another; every nuanced perspective is reduced to a viral soundbite.

Yet, when they meet face-to-face and start speaking, something changes: residents begin listening more than speaking, and people begin testing ideas together. Misunderstandings fade, and trust begins to form. By the end of their discussion, they have not only softened their hostility, but discovered actionable plans that benefit everyone.

This is the kind of meaningful discourse our society desperately needs. Yet our digital platforms—designed primarily for maximizing engagement through provocative content—have pulled us away from these core community endeavours.

As a constructive path forward, we introduce the idea of *conversation networks* as a basis for civic communication infrastructure that combines interoperable digital apps with the thoughtful integration of AI guided by human agency.

## The Gap in Our Digital Infrastructure

Community building relies on three forms of communication:

1. **Bridging**: Bringing people together across divides, helping to reduce polarization and foster understanding between fragmented groups.
2. **Listening**: Leaders—whether in organizations or communities—hear a wide range of authentic voices and perspectives, ensuring that everyone has a chance to be heard.
3. **Deliberation**: Collective reasoning, testing ideas, and generating actionable decisions.

What if we could harness the same elements that make social media powerful—its ease of use, habit-forming designs, and interconnected networks—to create something fundamentally different? Imagine scalable digital spaces designed for constructive communication: networks built around live, spoken conversations, rather than divisive content. These "**conversation**

---

[1] Massachusetts Institute of Technology & Cortico, dkroy@mit.edu
[2] Harvard Law School, lessig@law.harvard.edu
[3] Mozilla Foundation, Project Liberty Institute, and supported by funding from Omidyar Network, audreyt@audreyt.org

**networks**" offer a promising path to rebuild our fragmented social fabric, reduce polarization, and strengthen democracy.

By redesigning key elements of social media, including AI for content analysis, intuitively-designed apps, and digital networks, and combining them with thoughtful investment, human-centered training, and shared standards, we can catalyze a new ecosystem. With the right support, conversation networks could transform how we engage with one another, and meet the urgent needs of communities worldwide.

## What Are Conversation Networks?

In a conversation network, **the core "content" are excerpts from recorded group conversations**, not provocative social media posts or status updates. These recorded dialogues can be held in person or on digital platforms designed to encourage thoughtful discussion, broad participation, and respectful exchange. With consent of participants, excerpts from these conversations can then be heard by others to form new connections beyond the original conversation.

For example, if two groups in a community have become mutually polarized and distrustful, exposure to each other in performative public spaces such as social media and open-mic town hall meetings can exacerbate divisions. Instead, members of each group could hold recorded conversations amongst themselves that are structured and facilitated to surface nuanced perspectives. Excerpts from the conversation recordings can then be shared across groups, enabling each to hear voices and authentic sentiments from the other. Digital infrastructure to make such exchanges easy and scalable are what we mean by conversation networks—networks formed by the sharing of content that originates from live spoken conversation.

The quality of content in this approach is shaped by the conversational context; if the group conversation is facilitated and well-structured, the content emanating from it and flowing through the conversation network will be high quality. Instead of rewarding the loudest or angriest voices, these networks can foster empathy, nuance, and the search for mutual understanding.

Underneath the hood, a suite of tools—from AI-assisted transcription, analysis, and summarization, to speech and video conferencing apps—can enable these conversations to spread. Platforms such as **Cortico** and tools from the **MIT Center for Constructive Communication** facilitate meaningful **dialogue**, **listening**, and **sensemaking**. Meanwhile, **Polis and Remesh** facilitate **large-scale AI-supported collaboration,** while, **Frankly**, and the **Stanford Online Deliberation Platform** support large-scale, structured video-based **deliberation**. By integrating the best of face-to-face and digital engagement, conversation networks aim to restore the social fabric worn thin by years of digital discord.

## Three Examples

**vTaiwan: Dialogue → Deliberation → Policy Formation**

vTaiwan[4]—a prototype of an open consultation process for society to engage in responsible discussion on national issues—exemplifies a structured approach to collaborative governance, combining dialogue, deliberation, and policy formation. The process begins with broad conversations in weekly meetups, where stakeholders and citizens identify key issues and perspectives. Next, participants use Polis, an innovative AI-supported deliberation tool, to engage in online discussions that uncover areas of consensus and divergence. Insights from Polis then feed back into smaller, facilitated conversations, where participants refine collective priorities and produce actionable steps. Government officials ultimately translate these insights into policies rooted in public input and shared understanding.

Although vTaiwan distinguishes itself as the first large-scale deployment of Polis, facilitated conversations are equally vital. Pre-Polis discussions help generate high-quality, diverse opinion statements; post-Polis multi-stakeholder dialogues build on "bridging" statements[5] identified by the platform, transforming them into practical policy recommendations. With enhanced digital infrastructure, vTaiwan could become a conversation network that integrates in-person recorded discussions with Polis so that all people involved in the deliberations could hear and be heard beyond their small group discussions. We envision infrastructure enabling excerpts from conversations to be linked to inputs and outputs of Polis, and participants provided with intuitive digital apps to listen to the voices of others. The result would be an end-to-end model for inclusive, data-informed governance that builds understanding between people.

**Newark Youth Voices: Dialogue → Sensemaking → Advocacy**

Newark Youth Voices leverages dialogue and technology to empower young people in shaping their community's future. The process begins with dialogue, where youth participate in in-person, recorded conversations, sharing their experiences and perspectives on issues affecting their lives. These recordings are then analyzed using Cortico's platform,[6] which facilitates sensemaking by identifying key themes, patterns, and insights from the conversations. The conversations are designed and led by youth leaders, who also drive the sensemaking work. This process ensures that diverse voices are meaningfully represented. The insights are then mobilized to drive youth advocacy, equipping the Newark Opportunity Youth Network (OYN) team with the evidence and narratives they need to train teachers, engage decision-makers, propose solutions, and advocate for policies that reflect their collective priorities and lived experiences. The primary output from the Cortico platform is a "voice portal"[7] which provides public access to excerpts from group discussions. If Polis and Cortico were interoperable, selected excerpts from Cortico could be imported into Polis to seed deliberations—which could advance the advocacy work of OYN.

---

[4] Details about the vTaiwan platform are available at https://info.vtaiwan.tw/
[5] For more on bridging systems, see the "Prosocial Media" paper https://arxiv.org/abs/2502.10834
[6] Details about the Cortico platform are available at https://cortico.ai/platform/
[7] The portal for this project can be found at https://newarkyouthvoices.portal.fora.io/

**Deliberations.US: Information → Deliberation → Understanding**

Deliberations.US builds on Frankly's video-based deliberation platform to facilitate discussion about core issues of democracy. Participants view short (<5m) videos about topics of American democracy, including the electoral college, and money in politics. The content for these videos has been drawn from advisory panels, assuring a balanced presentation of the issues. Participants are assigned into small, video-based discussion groups, balanced based on demographics relevant to the deliberation. The platform moves them through a deliberation. Having measured attitudes at the start, the platform again measures attitudes after the deliberation, tracking how understanding has developed across demographics. If the discussion groups in Frankly are recorded, and Frankly were interoperable with Cortico, then the sensemaking and portal output features of Cortico would become easily available to Deliberations.US. This could enable an organized display of excerpts from deliberations associated with recommendations and outputs of the deliberation as a way to increase transparency and trustworthiness.

## A Fragmented Ecosystem

All three examples illustrate the potential of structured conversation networks, deliberation systems, and sensemaking analytics. They also highlight a central challenge: the tools exist, but they often remain fragmented. One process stage may rely on cutting-edge AI, while another stage involves analog conversations, manual sorting through transcripts, or scattered collaboration on social media. For example, Cortico's sensemaking features could in theory be used in vTaiwan to help translate pre-Polis discussions into input for Polis, or to help organize and publicly present output from Frankly, but the three platforms do not currently interoperate—there is no plug-and-play data standard to connect the digital systems.

When many people or organizations want to collaborate, it's vital that their digital tools fit together seamlessly through **open standards.**[8] This 'plug-and-play' approach enables several key features:

1. **Reduced Friction**: Communities can adopt multiple tools without painstaking integrations.
2. **Broader Adoption**: Lower technical barriers make it easier for diverse groups to try conversation networks.
3. **Shared Learning**: When data and insights move smoothly across platforms, best practices emerge more rapidly.

Think of it like building with LEGO blocks: if every block snaps together, anyone can assemble a conversation network tailored to their community's needs.

---

[8] Open standards are publicly available, consensus-driven specifications that define how digital tools and systems communicate and exchange data. They allow different software, hardware, and networks to work together seamlessly, free from proprietary restrictions or vendor lock-in.

## Fostering Shared Vocabulary, Concepts, and Skills

When different conversation network tools work together, it does not just make tool use easier—it helps people develop a common language for how they talk, listen, and make decisions. As more communities adopt interoperable tools, they will naturally share vocabulary and methods, making it easier for everyone to learn from each other. For example, terms including "conversation guide" / "discussion guide" and "sensemaking" are used in overlapping ways by dialogue and deliberation practitioners, yet differences in what the terms connote impede the sharing of best practices. A standard for tools that support these processes would lead to greater conceptual alignment and foster easier learning across practitioners.

While our emphasis here has been on digital tools, it is essential that people develop new communication skills and habits to effectively use these tools to foster constructive communication. Using these tools requires training, a commitment to community engagement, active listening, collaborative problem-solving, and thoughtful deliberation. It also needs a culture of accountability, transparency, inclusivity, and adaptability. By investing in human capacity-building—such as training for facilitation and deliberation practices—alongside technological innovation, we can ensure that these tools serve as enablers of community building.

The combination of conversation-centric tools and methods has the potential for wide adoption, spreading into schools, workplaces, and neighbourhoods. Grassroots conversation networks could emerge, rebuilding the trust and agency that have eroded over decades under the pressure of top-down media, social media, and national-level political polarization and fragmentation. Such networks would not only strengthen communities but also help lay the foundation for a more resilient society.

## The Promise and Peril of Automation

As we envision a cohesive ecosystem of tools to support conversation networks, we must also grapple with the role of automation in conversation networks, ensuring that its application actively strengthens, rather than undermines, community connections.

Large language models (LLMs) are already showing transformative potential in shaping how we engage with information and public discourse. AI-driven summarization of conversations (e.g., fathom.ai, otter.ai) and AI-led interviews to understand public opinion (e.g., talktothecity.org) are being actively used today, while AI-mediated deliberation tools, like the Habermas machine[9] remain a promising focus of ongoing research.

The increasing integration of automated processes in bridging, listening, and deliberation, reflects a broader trend: the growing reliance on technology to predict and, in some cases, replace human participation. Imagine taking this trend to its logical extreme: every citizen could be represented by an AI avatar,[10] continuously engaging in community discourse on their behalf—what we might call an "Avatar State." Would such a development fulfill John Dewey's vision of an inclusive participatory democracy, in which all citizens—given the right educational and institutional support—can actively engage in self-governance? Would it address the challenges Walter Lippmann identified as significant obstacles (that most people lack the expertise, time, or interest to participate effectively in governance)? Or would it veer toward a techno-autocracy, where the essence of human agency is handed over to those who control the AI?

---

[9] Michael Henry Tessler et al., "AI Can Help Humans Find Common Ground in Democratic Deliberation," *Science* 386, no. 6719 (October 18, 2024), https://doi.org/10.1126/science.adq2852.

In designing our future, we must consider the ideal roles AI can and should play—and, just as importantly, the roles that only people can fulfill.

## AI's Assistive Potential in Strengthening Conversation Networks

AI technologies could displace human action or they could complement, by assisting human bridging, listening, deliberation. We believe they have enormous potential to be transformative and constructive in strengthening conversation networks when applied in an assistive capacity:

1. **Expand Participation**: Offering supportive structures and learning resources can broaden engagement, inviting people from any background or level of expertise to take part in a meaningful way.
2. **Spread Effective Practices**: AI can help disseminate effective practices, allowing communities to learn from each other's successes and build upon established methods.
3. **Reveal Blind Spots**: AI can uncover overlooked perspectives and issues, shedding light on concerns that might otherwise remain hidden.
4. **Automate Repetitive Tasks**: Finally, AI can handle time-consuming "spade work," freeing up time for human participants to focus on contextualized decision-making and authentic human-to-human collaboration.

In democratic processes such as citizen assemblies, AI should not come between people in ways that diminish or cut off direct human-human conversation. Instead, AI should be used to ensure members engage fully with one another, promoting genuine human interaction and deliberation without technological mediation. Learning how to work through differences of perspective and opinion is an integral part of the deliberative process that should be supported and protected. AI could help enable that.

Bridging networks,[11] such as scaled community networks on U.S. campuses, present another opportunity. In these networks, AI could support logistical and analytical tasks, while humans perform all key functions such as facilitation and overseeing sensemaking, fostering authentic connections and trust.

---

[10] A recent study demonstrates that AI simulations of individual people, based on interviewing them, can accurately predict their attitudes and behaviours. See: Joon Sung Park et al., "Generative Agent Simulations of 1,000 People" (arXiv, November 15, 2024), https://doi.org/10.48550/arXiv.2411.10109.

[11] A conversation network that bridges groups of people by enabling them to hear cross-cutting perspectives, experiences, and opinions.

## Preserving Human Agency and Accountability

Although AI can lighten the burdens of analysis and coordination, people must remain at the heart of conversation networks. We need to build *civic muscle*—the skills, habits, and capacities required for active participation—so communities can exercise agency and take ownership of their roles in governance. It is critical that humans, not AI tools, own their actions and take credit as well as responsibility for decisions and outcomes shaped by AI.

AI used in the context of conversation networks must therefore operate within guardrails, refusing tasks that require making subjective judgments about what other people feel or want. Instead, AI should be focused on more objective tasks—such as organizing data or highlighting patterns—while leaving interpretation and moral responsibility to human beings. Performance audits, in which even non-experts critically evaluate how AI systems are functioning, are equally important to keep under human control.

Maintaining decentralized control in the creation and shaping of conversation networks is also essential, with a particular emphasis on fostering community-controlled AI to prevent power from concentrating in the hands of a few. Lastly, humans must actively practice interpersonal connection, ensuring that technologies mediating our interactions enhance and deepen relationships, rather than erode them.

By placing human agency, trust, and accountability at the heart of how we use AI, we can ensure that AI remains a catalyst for authentic conversation and collective action—rather than a replacement for it.

## Conclusion

The need for meaningful civic discourse has never been greater, and the tools to support it are within reach. Conversation networks, supported by thoughtful integration of AI and guided by human agency, offer a path forward. By addressing current gaps, fostering interoperability, and investing in both technology and the people who use it, we can create scalable systems that empower communities, build trust, and strengthen societies worldwide.

But technology alone is no panacea. We must invest in human skill-building at every level—teaching people how to hold nuanced conversations, make sense of what others mean, weigh evidence critically, and co-create solutions. At the same time, we must establish guiding design principles, open standards, and ethical safeguards so that AI bolsters—rather than undermines—our democratic goals.

The stakes could not be higher. Polarization and mistrust will continue to unravel our social fabric unless we choose to act. We have at our disposal the tools and know-how to transform our digital landscape, but the key question remains: are we willing to invest in both human and technological capacities to fulfill this vision?

The promise of conversation networks is more than just a hope for better social media; it is an opportunity to reclaim our collective agency and renew our sense of community. If we seize this moment, we can catalyze nothing less than the regeneration of our shared civic life.